\documentclass[sigconf,screen,authorversion]{acmart}

\usepackage[utf8]{inputenc}        % ✓ Encoding of this TeX source file
\usepackage[nameinlink]{cleveref}  % ✓ Clever references
\usepackage{listings}              % ✓ Formatted code blocks

\copyrightyear{2024}
\acmYear{2024}
\setcopyright{rightsretained}
\acmConference[ITiCSE 2024]{Proceedings of the 2024 Innovation and Technology in Computer Science Education V. 1}{July 8--10, 2024}{Milan, Italy}
\acmBooktitle{Proceedings of the 2024 Innovation and Technology in Computer Science Education V. 1 (ITiCSE 2024), July 8--10, 2024, Milan, Italy}
\acmDOI{10.1145/3649217.3653642}
\acmISBN{979-8-4007-0600-4/24/07} 

\begin{document}

\title{Research and Practice of Delivering Tabletop Exercises}

\author{Jan Vykopal}
\orcid{0000-0002-3425-0951}
\affiliation{
    \institution{Masaryk University}
    \department{Faculty of Informatics}
    \city{Brno}
    \country{Czech Republic}
}
\email{vykopal@fi.muni.cz}

\author{Pavel Čeleda}
\orcid{0000-0002-3338-2856}
\affiliation{
    \institution{Masaryk University}
    \department{Faculty of Informatics}
    \city{Brno}
    \country{Czech Republic}
}
\email{celeda@fi.muni.cz}

\author{Valdemar Švábenský}
\orcid{0000-0001-8546-280X}
\affiliation{
    \institution{Masaryk University}
    \department{Faculty of Informatics}
    \city{Brno}
    \country{Czech Republic}
}
\email{valdemar@mail.muni.cz}

\author{Martin Hofbauer}
\orcid{0009-0005-3998-9164}
\affiliation{
    \institution{Masaryk University}
    \department{Faculty of Informatics}
    \city{Brno}
    \country{Czech Republic}
}
\email{hofbauer@fi.muni.cz}

\author{Martin Horák}
\orcid{0000-0002-1835-6465}
\affiliation{
    \institution{Masaryk University}
    \department{Faculty of Informatics}
    \city{Brno}
    \country{Czech Republic}
}
\email{horak.martin@fi.muni.cz}

%% Define a concise list of authors' names for page headers.
%\renewcommand{\shortauthors}{Anonymous author(s)}

\begin{abstract}
Tabletop exercises are used to train personnel in the efficient mitigation and resolution of incidents. They are applied in practice to support the preparedness of organizations and to highlight inefficient processes. Since tabletop exercises train competencies required in the workplace, they have been introduced into computing courses at universities as an innovation, especially within cybersecurity curricula. To help computing educators adopt this innovative method, we survey academic publications that deal with tabletop exercises. From \papers\ papers we identified and examined, we selected \papersselected\ papers for a detailed review. The results show that the existing research deals predominantly with exercises that follow a linear format and exercises that do not systematically collect data about trainees' learning. Computing education researchers can investigate novel approaches to instruction and assessment in the context of tabletop exercises to maximize the impact of this teaching method. Due to the relatively low number of published papers, the potential for future research is immense. Our review provides researchers, tool developers, and educators with an orientation in the area, a synthesis of trends, and implications for further work. 
\end{abstract}

%% The code below is generated by http://dl.acm.org/ccs.cfm.

%% Pick words that accurately describe the work. Separate with commas for PDF and semi-colons for the ACM web form.

\newcommand{\papers}{140}
\newcommand{\datesearchended}{July 19, 2023}
\newcommand{\papersselected}{14}

\maketitle

\section{Introduction}
\label{sec:introduction}

The Computing Curricula 2020 report (CC2020)~\cite{cc2020} responds to the need for graduates who are effective in their work roles and tasks. It promotes using \emph{competencies} instead of knowledge to describe computing curricula.
Competency augments knowledge with its skilled application motivated by the purpose of accomplishing a task. According to CC2020, the knowledge of subject matter and related skills are as important as analytical and critical thinking, collaboration and teamwork, and dispositions such as being responsible and flexible or having self-confidence and self-control. 

Computing educators research and use various forms of experiential learning to prepare students for their future careers, such as simulations, team projects, or industry internships. We highlight \emph{tabletop exercises} (TTX) as an efficient method for gaining competencies relevant to work roles and tasks centered around analysis, decision making, and communication with various parties. In particular, these exercises are relevant for cybersecurity and IT governance courses~\cite{enisa-guide}, addressing incident response methodologies outlined in frameworks like COBIT (Control OBjectives for Information and related Technology) and ITIL (Information Technology Infrastructure Library). While tabletop exercises are common in professional settings, they have not been widely used in computing courses. We believe that these exercises have a great potential for innovating teaching practice in higher education institutions.

This paper examines the development and state of the art of TTX as presented at various academic venues. The core contribution is a systematic review of research papers to understand state of the art. Our work is useful to various target groups. For educators, it shows examples of approaches and exercises and presents practical recommendations. For researchers, it provides an overview of methods used in exercises and implications for further research. Finally, developers of educational tools can be inspired by the existing tools and implications for future work.

\section{Background}
\label{sec:inject}

A tabletop exercise is a form of a teaching activity aimed at training teams in responding to crisis situations~\cite{nist-guide}. It involves simulating a crisis within the context of business operations in an organization, such as a ransomware attack on the company infrastructure or exfiltration of sensitive information. The team members (exercise participants) assume different roles in the organization, such as chief security officer or cybersecurity incident responder~\cite{Angafor2023}. During the exercise, they discuss how to effectively respond to the crises while adhering to processes and regulations. Instructors facilitate these discussions and provide a debriefing after the TTX. The exercise usually lasts a few hours or days at most.

TTXs are driven by \textit{injects}, pre-scripted messages, which can take the form of an e-mail or a news article, provided to trainees during the exercise. The purpose of injects is to advance the exercise and stimulate further actions and discussions. For instance, injects can notify teams of a data breach in their organization, requiring them to respond appropriately~\cite{nist-guide}.

TTXs focus on communication, coordination, and collaboration~\cite{Angafor2023}, not particular technical skills as it is the case during hands-on training in an emulated IT environment (such as a cybersecurity lab~\cite{cyber-arena} or cyber range~\cite{yamin2020}). The nature of TTXs allows to conduct them using pen and paper or simple online office applications (such as Microsoft Forms), making TTXs relatively cost-effective. However, assessing the trainees is not automated -- it requires instructors' manual effort, which is highly time-consuming. It may take days or weeks until the trainees receive feedback from instructors, which lowers the effectiveness of the exercise.

TTXs share certain traits with some other forms of active learning~\cite{Sanders2017} but also have unique characteristics. For instance, course projects in software development or Process oriented guided inquiry learning (POGIL)~\cite{pogil2012} also involve student teams that work together. In contrast, TTXs do not feature a clearly specified and structured assignment. Moreover, student teams in a TTX are not guided by instructors but decide on their own about the current priorities and tasks to complete. Last, TTXs intentionally overwhelm trainees with numerous pieces of information and inputs to simulate a stressful emergency situation.

TTXs in the cybersecurity domain can be conducted either independently or as a part of complex exercises involving technical skills (denoted as cyber defense exercises or red vs. blue team exercises). Locked Shields~\cite{locked-shields} and Cyber Europe~\cite{cyber-europe} are exercises combining TTXs with technical training. Both are centered around a background story resembling a recent real crisis or attack campaign~\cite{Vykopal2017,Ostby2019}. Participants representing one organization or country are assigned different roles and divided into teams. While some teams exercise mainly technical skills in an emulated IT environment, others are engaged in decision-making processes, standard operational procedures, or communication in a local and international context. Locked Shields is the largest global defense exercise organized by the NATO Cooperative Cyber Defence Centre of Excellence in Tallinn, Estonia, since 2010.
Cyber Europe is a series of pan-European exercises developed for IT security, business continuity, and crisis management teams by European Union Agency for Cybersecurity (ENISA), since 2010.

Diverse commercial solutions are available to address various aspects of emergency preparation. Some vendors, such as PreparedEx \cite{PreparedEx} or Emergency Solutions International~\cite{ESI}, provide comprehensive TTXs for crises, training emergency responders, and highlighting inefficiencies in processes and inadequate capabilities. Numerous commercial services (e.g., \cite{CrowdStrike, Microminder, RedLegg, CyberSecOp}) focus on validating incident response and business continuity plans in IT operations or cybersecurity and are often part of broader services portfolios. Additionally, there are platform-as-a-service solutions (namely~\cite{Conducttr, Cinten, Eeedo, Avalias}), all focused on crisis management in general. Furthermore, specialized platforms, such as Cyber Crisis Simulator~\cite{ImmersiveLabs} or open-source OpenEx~\cite{openex}, are designed specifically for IT and cybersecurity crisis management.

National or international authorities, such as NIST~\cite{nist-guide}, FEMA~\cite{HSEEP}, ENISA~\cite{enisa-guide}, or ISO~\cite{ISO22398} published standards and guides on complex exercises, which also include TTXs or can be applied to them.

To conclude, a TTX is an established training method used in practice, yet mostly outside university settings. Our review maps the state of the art based on academic publications on this topic.

\section{Previous Literature Reviews}
\label{sec:related-work}

In the academic literature, there is one review of TTXs in the cybersecurity domain and three papers in the subject area of healthcare.

\citet{Angafor2020} reviewed academic and commercial product literature on TTXs used for training cybersecurity incident response teams. The scope of our review is wider; we searched for papers in broad subject areas of computer science and engineering. Also, our review captures recent research and trends since 2020 when the other review was published. The overlap of this and their study is only two papers (namely P4 and P7, see details in~\Cref{table:papers}). 

Mahdi et al.~\cite{mahdi2023} reviewed disaster preparation exercises conducted by academic healthcare institutions. Based on the reviewed literature, the authors concluded that TTXs are the easiest to organize, conduct, and evaluate, while also useful in evaluating emergency response protocols and their subsequent improvement.

Evans~\cite{Evans2019} surveyed healthcare literature for using TTXs in nursing education. The opportunity to identify knowledge gaps, as well as knowledge gain, was reported. The author also proposed a list of considerations for exercise development.

Finally, Fr{\'e}geau et al.~\cite{Fregeaue2020} published a scoping protocol for a review that maps the uses of TTXs in healthcare. However, the review itself has not been published yet.

% ==================== Section start ====================
\section{Method of Conducting the Review}  

\begin{figure*}[!ht]
    \centering
    \includegraphics[width=0.95\textwidth]{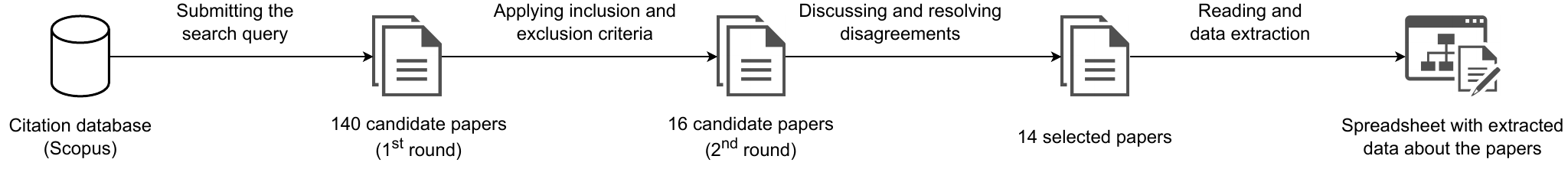}
    \caption{Steps of the systematic literature review and the number of processed papers.}
    \Description{Diagram showing the linear progress of the steps of the literature review.}
    \label{figure:SLR-process}
\end{figure*}

We follow the guidelines for a systematic literature review (SLR)~\cite{kitchenham2007} and a systematic mapping study~\cite{Petersen2008, Petersen2015}. This section presents the SLR protocol, which specifies the research questions, search process (see~\Cref{figure:SLR-process}), and criteria for including the discovered papers.

\begin{figure*}[!ht]
    \centering
{\small
\begin{verbatim}
TITLE-ABS-KEY( "table* exercise" OR "communication exercise" ) AND ( LIMIT-TO ( SUBJAREA, "COMP" ) OR LIMIT-TO ( SUBJAREA, "ENGI" ) ) 
\end{verbatim}}
    \caption{The query for the automated paper search for papers in Scopus~\cite{scopus}.} %TITLE-ABS-KEY searches in the title, abstract, and keywords.}
    \Description{Text-based search query input in the Scopus database.}
    \label{figure:search-query}
\end{figure*}

\subsection{Research Questions}
\label{subsec:rqs}

Our literature review examines five research questions:
\begin{enumerate}
    \item \textit{What formats of tabletop exercises are used?}\\
    Namely how the exercises are prepared, delivered and what tools are applied.
    
    \item \textit{Who are the participants of the exercises?}\\
    What are the target groups of the exercises (trainees) and by whom are they organized (instructors, designers)?
    
    \item \textit{How are the exercises developed, assessed, and evaluated?}\\
    We will examine methods for exercise preparation, assessment of trainees, and evaluation of the exercise itself.
    
    \item \textit{How are the research results applied in practice?}\\
    Do the publications provide any supplementary materials or artifacts for other educators?
    
    \item \textit{What are the future research directions and challenges?}\\
    What is the ongoing and long-term goal?
\end{enumerate}

\subsection{Paper Search and Automated Filtering}
\label{subsec:search}
We used the Scopus citation database of peer-reviewed literature~\citep{scopus}. Scopus indexes a representative portion of the databases of individual publishers, including ACM Digital Library, IEEExplore, SpringerLink, or Elsevier ScienceDirect. We did not use Google Scholar since it additionally indexes many non-peer-reviewed papers.

When constructing the search query, we focused on the subject area of computer science and engineering and used ``tabletop exercise'' as the primary keyword. We also sought not to miss potential alternative names for tabletop exercises, such as ``simulation'' or ``communication exercises''. After several pilot searches, we concluded with the query presented in~\Cref{figure:search-query}. The asterisk is a wildcard to match both adjectives ``table top'' and ``table-top''. However, the wildcard is not used for the stem of the word ``communication'' because the pilot search yielded numerous false positives. Similarly, we omitted ``simulation'' matching a huge number of papers unrelated to education and training. The search was case-insensitive.

We started the search on May 11, 2023, and then subscribed to Scopus e-mail notifications that informed us about newly indexed papers, which we gradually added to the candidate set. We stopped adding new candidates on \datesearchended. In total, we found \papers\ candidate papers.

\subsection{Inclusion and Exclusion Criteria}

After multiple pilot tests, we set ten criteria. 
\begin{enumerate}
\item Papers must deal with TTXs in IT or operational technology (OT) operations or security. 
\item Papers describing exercises and tools in other domains, such as transportation or utilities, without considering cybersecurity or IT/OT operations aspects, are excluded.
\item We include papers written in English with \emph{full text available}\footnote{We consider a paper to be available if we can access it using tens of licensed electronic resources our institution has access to, or if the full text is freely available online.}. We do \emph{not set any page limit} to include short papers, which report ongoing and future work at the time of their publication.
\item Other types of documents yielded by the automated search, such as conference reviews or records, are excluded.
\item Papers must describe \emph{exercising or supporting the exercise of a complex process or standard operation procedures in a safe, simulated environment}. For instance, this includes experience reports with lessons learned or descriptions of tools supporting the preparation, execution, and analysis of TTXs.
\item Papers that describe only exercising essential communication and writing skills (e.g., technical writing, proposal paper, presentation and its delivery) are excluded.
\item Papers must report on an exercise involving \emph{teams or groups, not only individuals}.
\item The paper must support an \emph{educational goal}, for example, to train or assess participants, or help instructors conduct the exercise, or understand the learning processes.
\item We exclude papers that use TTXs solely as an evaluation tool in non-educational settings, such as for testing of software or IT systems.
\item \emph{Generic methodologies} applied or applicable to TTXs in IT and OT operations and security are included.
\end{enumerate}

Two authors of this paper each screened all \papers\ candidate papers and applied these criteria independently. The authors followed a simple algorithm for screening the paper content~\cite{Svabensky2022}:
\begin{center}
\begin{lstlisting}
    |\textbf{for}| each paper in the candidate set:
        read the title and abstract
        decide for inclusion or exclusion
        |\textbf{if}| decision cannot be made:
            read the introduction and conclusion
            decide for inclusion or exclusion
            |\textbf{if}| decision cannot be made:
                skim-read the rest of the paper
                decide for inclusion or exclusion
\end{lstlisting}
\end{center}

\subsection{Selecting Papers for Review}

After both authors finished their reading, they compared candidate papers they identified after applying the inclusion and exclusion criteria. If both authors decided to include the paper, it was selected for review. If both decided not to include the paper, it was excluded. In case of a disagreement, the authors discussed their views and agreed on the final decision.

In the first round, the authors agreed that out of 140 candidate papers, nine (6.4\%) were fitting the selection criteria. The authors had opposite opinions on seven (5\%) papers. Out of the 16 papers, which passed the first round, the authors selected \papersselected\ papers and rejected two papers in the second selection round.
Their inter-rater agreement~\cite{krippendorff2004reliability} measured by Krippendorff's $\alpha$ for nominal data was 0.69, which is substantial. The coefficient was calculated using the Python NLTK module~\citep{nltk}.

\section{Results}
\label{sec:results}

This section presents data extracted from reading the full texts of the \papersselected\ papers to answer the research questions posed in \Cref{subsec:rqs}. \Cref{table:papers} summarizes the goals and types of the reviewed papers. The summary of the papers is provided in \Cref{sec:discusssion}. Further, we refer to the selected papers using arbitrary numbered identifiers $P_x$.

\begin{table*}[!ht]
\caption{Overview of the \papersselected\ reviewed papers ordered by the year of publication. Paper Type: SW = software tool, E = exercise instance (run of a particular exercise), F = exercise format (method for conducting the exercise), O = other topic (see Goal).}
\vspace{-1em}
\centering
\renewcommand*{\arraystretch}{1.2}
\begin{tabular}{llcl}
\textbf{Paper ID} & \textbf{Year} & \textbf{Paper Type} & \textbf{Goal of the Paper} \\ %& \textbf{Country} & \textbf{Citations} \\
\hline
P1~\cite{Marshall2009} & 2009 & SW, F & Describe a tool for planning complex functional and tabletop exercises \\ %& USA & 3 \\
P2~\cite{Ottis2014} & 2014 & F, E & Describe a format of lightweight exercises \\ %& Estonia & 10 \\
P3~\cite{Svante2014} &  2014 & O & Propose design ideas helping to incorporate more %``wicked'' 
problems having no obvious solution into TTXs \\ %& Sweden & 8 \\
P4~\cite{Line2015} & 2015 & E & Study challenges when performing TTXs \\ %& Norway % SINTEF & 11 \\
P5~\cite{Makrodimitris2015} & 2015 & O & Review of Hermes, MSB and ENISA exercise methodologies w.r.t. the ISO standard 22398 \\ %& Greece & 2 \\
P6~\cite{Janssen2015} & 2015 & O, E & Introduce a method for trainee assessment \\ %& Netherlands & 1 \\ 
P7~\cite{brilingaité2017} & 2017 & SW, F, E &  Present a web-based environment for conducting TTXs \\ %& Lithuania & 7 \\
P8~\cite{taniuchi2019} & 2019 & E, O & Study errors made by TTX facilitators when interacting with trainees \\ %& Japan & 1 \\
P9~\cite{Kopustinskas2020} & 2020 & E & Report takeaways from a particular exercise \\%& Italy & 0 \\
P10~\cite{angaforg2020} & 2020 & O & Investigate the current cybersecurity skills gap and how TTXs can fill it \\ %& UK & 4 \\
P11~\cite{yuitaka2022} & 2022 & F, E & Describe three exercises developed by the authors \\ %& Japan & 0 \\
P12~\cite{Angafor2023} & 2023 & O, E & Report experience from a virtual incident response tabletop exercise \\ %& UK & 0 \\
P13~\cite{skytterholm2023} & 2023 & O, F & Investigate the characteristics of a realistic and expedient exercise scenario \\ %& Norway & 0 \\
P14~\cite{Zacharis2023} & 2023 & SW, F & Apply machine learning to unstructured information sources to generate exercise content \\ %& Greece & 0 \\
\hline
\end{tabular}
\label{table:papers}
\end{table*}

\subsection{RQ1: What Formats of Exercises Are Used?}
\label{sec:results-RQ1}

From seven papers (namely P1, P2, P4, P7, P12, P13, P14) out of eight that contain enough information about the exercise format, we see that the exercises are designed as a series of injects (events, problems, or situations, see \Cref{sec:inject}). These injects form a scenario that is unknown to the exercise participants beforehand. The injects are provided by exercise facilitators to participants who discuss appropriate actions, processes, or best practices within their team or with all participants. 

The eighth paper (P11) reports two kinds of exercises that extend this common format. One TTX uses cards that drive the exercise itself (attack cards) or stimulate participants to think about several options (action cards, situation awareness cards, and information-sharing cards). Another TTX in P11 tasked participants to create a scheme depicting information and workflows within an organization affected by an incident.

Only four papers mention the use of any software tool during the exercise. 
P7 presents a web application for delivering a TTX involving various roles according to a scenario defined by the facilitator. Participants use a graphical interface to respond to presented injects. They can submit a short description of how they would solve the inject, inform other roles about it, or delegate the resolution to other roles.
P11 mentions an exercise where participants use software to interact with virtual participants. Participants choose a response to presented injects from a limited set of predefined actions.
P12 reports a remote exercise facilitated through
Microsoft Teams web conferencing application.
P1, P6, and P14 use a software tool before or after the actual exercise; see~\Cref {sec:results-RQ3}.

The exercises last from several hours (P2, P13, P14) through one day (P4, P7, P8, P9, P11, P12) to a few days (P6).

\subsection{RQ2: Who Are the Exercise Participants?}

Trainees come from diverse sectors. The most frequent were critical infrastructure organizations, such as energy distribution operators (P4), a water management centre (P6), industrial control systems stakeholders (P8, P11), or oil and gas suppliers (P9, P13). 
Two exercises were carried out for university students (P2, P7).
One exercise was conducted for a large law enforcement organization (P12). The number of trainees ranged from 20 to 108. 

Exercises were designed by representatives of national or transnational authorities (P4, P9) or academic staff (P2, P8, P11, P12). The type of the organizing entity determines the target group and its diversity (employees of one organization, multiple organizations in a single country or multiple organizations from multiple countries). 

\subsection{RQ3: How Are the Exercises Developed, Assessed, and Evaluated?} \label{sec:results-RQ3}

\subsubsection*{Development}

There is no prevailing trend in the process of exercise preparation in the reviewed papers, even though guidelines~\cite{HSEEP,enisa-guide,nist-guide} and standards~\cite{ISO22398} had already been published.

The exercises in P4 and P7 were designed and conducted using guidelines from NIST~\cite{nist-guide} and ENISA~\cite{enisa-guide}, respectively.

P1 introduces a web-based collaborative tool for designers of cybersecurity TTXs, enabling scenario creation based on TTX objectives following a national guideline for exercise development~\cite{HSEEP}. It supports various user roles like Scenario Designer, Subject Matter Expert, or Observer, in producing a detailed list of scenario events.

P5 studies the compliance of three guidelines from the European authorities for designing and conducting exercises with the international standard ISO~22398~\cite{ISO22398}.

P13 investigates the characteristics of a realistic and expedient scenario in the field of industrial control systems. The paper lists 21 criteria and eight exercise topics based on these scenarios.

P3 argues that current tabletop exercises do not accurately reflect the reality of uncertain and complex problems that do not have obvious solutions. % (so-called ``wicked'' problems). 
It presents three design ideas for designing more efficient exercises. First, the focus should be on unsolved problems that the participants themselves come up with. Second, problems should be tamed during the exercise by the participants instead of during the planning phase by the designer. Third, the participants should use existing plans and experience from previous emergencies to resolve the problems in collaboration.

P14 employs machine learning to generate scenarios of cybersecurity exercises. First, a corpus of publicly available articles about cybersecurity incidents is created and annotated to detect threat actors, incidents, and victims. Then, an exercise designer provides inputs such as a keyword or a sector for generating a graph of an incident for a created exercise. The graph is then enhanced using information mined from existing cybersecurity taxonomies. Finally, the graph is transformed to text using GPT-2.

\subsubsection*{Assessment}

The assessment of exercise participants (trainees and/or facilitators) is addressed only in three papers.

P6 proposes a method and a tool for structured assessment of trainees. First, observable behavior in exercise subgoals must be defined in the exercise preparation. During the exercise, human observers record five performance aspects (timeliness, accuracy, relevance, completeness, and cost-effectiveness) in four phases of the OODA loop (Observe, Orient, Decide, Act)~\cite{ooda}, and the tool produces a score capturing trainees' performance.

Exercises reported in P4 include unstructured assessment of
trainees’ actions by the observers and facilitators after the exercise.

P8 studies errors made by exercise facilitators during their interactions with groups of trainees.
P8 recognizes the error of \emph{commission} and the error of \emph{omission}, which can negatively affect the progress of a trainee group in the exercise.

\subsubsection*{Evaluation}

Six papers addressed the evaluation of the exercise itself. None reported the use of specific qualitative or quantitative research methods.

In P4 and P11, the evaluation is conducted as a reflection of trainees and a discussion with the exercise designers and facilitators after the exercise.

In the exercise described in P2, teams of trainees compete against themselves in opposing roles of attackers and defenders. The feedback session is then used to uncover the goals and motivations of the other role. Also, the instructors ask for immediate feedback on the exercise format.

P7 presents summary counts of trainees' actions in the exercise platform and messages sent during the exercise.

P9 reports the evaluation that has two phases. Firstly, the trainees shared comments about what they learned at the end of the exercise. Afterward, experts invited by the exercise organizers surveyed the trainees and evaluated the outputs produced during the exercise.

P12 discusses responses from a trainee survey, which relates the exercise content to incident response and disaster recovery plans of a particular organization where the exercise was held.

\subsection{RQ4: How Are the Results Applied?}

No reviewed paper refers to supplementary materials (such as software tool implementation or exercise scenario) that other educators can directly use.
In particular, P1, P6, P7, and P14 describe software tools, but the papers do not refer to any materials, such as software repositories or websites presenting the tools.
P13 only outlines eight scenarios as an inspiration for creating a new exercise.

A few papers distill recommendations and lessons learned from TTX preparation or delivery applicable to other exercises. P2 lists steps describing the preparation of a lightweight TTX and the most common problems encountered. P3 provides three design ideas for more realistic and efficient exercises (see \Cref{sec:results-RQ3}). P4 states recommendations for conducting exercises.

Finally, P5 shows that the methodology by ENISA~\cite{enisa-guide} is the most compliant with the ISO 22398 standard out of the three reviewed methodologies, yet not entirely.

\subsection{RQ5: What Are the Future Directions?}

The reviewed papers reported diverse future work. 
P2 suggests developing more detailed instructions, creating mock scenarios, and formalizing and publishing student feedback.
P3 advises the development of an online tool for running a TTX, evaluating the proposed design ideas, and studying how TTXs are used in incident recovery.
P4 recommends studying best practices and challenges of organizations conducting TTXs regularly and studying real-life incident responses to design more useful future exercises.
P6, which deals with assessment, plans to replace binary scoring with a four-point scale in the presented method.
Since P7 piloted the exercise with university students, future work includes conducting the exercise with professionals working in the target industry. In addition, it propounds the development of an assessment of trainee stress levels, which can then be used to adapt the exercise progression dynamically.
P8 mentions sharing the instances of errors among facilitators. While P9 simply recommends conducting more exercises, 
P11 suggests developing a method to demonstrate the risks due to failure to act. P12 plans to improve the presented exercise using lessons learned from its first run and develop a training program for staff at all organizational levels. P13 would like to test the proposed criteria for realistic and expedient scenarios in practice when deploying the proposed scenarios.
Finally, P14 lists several concrete directions to improve the tool for generating exercise content. 

% ==================== Section start ====================
\section{Discussion} 
\label{sec:discusssion}

This section provides takeaways from the review of the selected papers. It helps the reader understand the greater themes of the results presented in \Cref{sec:results}. The review limitations are also discussed.

\subsection{Summary of the Observed Trends}
\label{subsec:summary}

The topics in the identified papers differ widely. Most papers are reports from runs of a particular exercise or describe an exercise format. The typical TTX format is based on pieces of information gradually provided to the trainees by exercise facilitators. The trainees discuss and respond to the information in teams. The exercises are usually held for tens of trainees on site for several hours or days, with little support from dedicated software tools. 

The explicit learning phase usually happens at the end or after the exercise, when trainees reflect on their decisions with other participants or when the previously unknown parts of the scenario are disclosed to them. Other types of assessment or feedback are rare because the complexity and labor of the manual preparation of the exercises hinder instructors from focusing on trainee assessment and exercise evaluation. Out of only three papers (P4, P6, and P8) that addressed assessment, only one (P6) suggested a method that goes further than unstructured assessment by the observers and facilitators. Similarly, the evaluation of the exercise itself is also underdeveloped. No paper uses advanced feedback methods but trainees' reflections, discussions, or surveys.

The reviewed papers did not provide many actionable artifacts or supplementary materials (such as software supporting exercise preparation and delivery, exercise playbooks, scenarios, or checklists) for others considering conducting their own exercise. The absence of publicly available materials prevents educators from adopting the concept of tabletop exercises as a teaching method.

The current practice of conducting TTXs relies on manual preparation of exercise content (namely injects that form exercise scenarios), with no automation and limited future reusability. We believe this practice can be improved by leveraging large language models for semi-automated content generation (P14) or employing dedicated software for TTX preparation and delivery (see \Cref{sec:results-RQ1}).

Finally, future directions in the papers often mention evaluating proposed methods by field tests or running the exercise for more trainees or different target groups. These plans indicate that the described methods and exercises should be developed further and provide new research opportunities in computing education.

\subsection{Limitations} 

This review is limited by narrowing its scope by only using the Scopus citation database of the peer-reviewed literature. This decision was explained in Section~\ref{subsec:search}. Also, when using Scopus, we did not have to deal with duplicate records, which may occur when using multiple other databases.

The second limitation is keyword choice. Using other keywords may have led us to find different papers fitting the selection criteria. However, pilot searches with related terms, such as ``crisis exercise'' returned an excessive number of unrelated papers (see \Cref{subsec:search}). 

Finally, the data extraction was done manually, which may lead to misinterpretation of the presented information. To mitigate this risk, we performed cross-author discussion and validation.

\section{Conclusions and Future Directions}
\label{sec:conclusions}

While tabletop exercises have been conducted for over a decade, often by governmental and military agencies, the number of published papers has grown only in recent years. The organizing entities have developed their guidelines and summarized the best practices, but there is no widely used terminology, textbooks, or tools for the exercise development, delivery, or evaluation. This situation is one of the causes why TTXs are not yet widely used in computing education, even though they can help develop competencies and dispositions relevant to students' future careers as IT professionals.

To the best of our knowledge, this paper is the first systematic review of TTXs in the context of IT operations and security. We examined 140 papers and reviewed 14 of them in detail. The structured information extracted from the identified papers is published as supplementary material~\cite{ttx-dataset}, which also includes a review of 16 practical implementations that did not fit this paper's page limit.

This review implies several possible directions for future research and practice. First, novel formats of the exercises can be proposed to make them more realistic but still lightweight on resources and effort required for their development and delivery. For instance, the current discussion-based approach (``What would your team/organization do when a certain event happened?'') could be innovated by letting the trainees perform the actual simulated actions (such as asking another team/external party to do something) and observe whether they do what was expected by the exercise scenario based on training objectives. Another example is an investigation of opportunities and limits of remote or hybrid exercise that would enable more trainees to participate and ease traveling and logistics. 

The next direction is to start using dedicated software tools in the TTX preparation, delivery, and evaluation, such as INJECT Exercise Platform~\cite{Svabensky2024from}. Automated tools can substantially increase exercise scalability and lower the preparation effort. Not only can they enable the participation of more trainees (regardless of whether the TTX is delivered onsite or remotely), but they also reduce the workload of exercise designers and facilitators. Moreover, researchers may explore employing machine learning for trainee assessment or leverage large language models to generate exercise content. 

Next, developers can provide a dedicated platform for designing and conducting the exercise, which would assist the facilitators or even automate some of their tasks (such as playing the role of some simulated actors in the exercise). Moving the exercise from pen and paper or online forms and documents to the dedicated platform would enable the collection of data about trainees' interactions during the exercise and further analysis for the assessment of trainees and evaluation of the exercise. Finally, the exercise designers would strongly benefit from releasing and sharing tangible outputs, such as software tools, scenarios, or checklists.

\begin{acks}
This research was supported by the Open Calls for Security Research 2023--2029 (OPSEC) program granted by the Ministry of the Interior of the Czech Republic under No. VK01030007 -- \textit{Intelligent Tools for Planning, Conducting, and Evaluating Tabletop Exercises}.
\end{acks}

% Authors' names should be complete -- use full first names.
\bibliographystyle{ACM-Reference-Format}
\balance
\bibliography{references}

\end{document}